\documentclass{ifacconf}

\usepackage{graphicx}      
\usepackage{natbib}        
\usepackage{url}

\usepackage{amsmath,amssymb}
\usepackage{color,svgcolor}

\newtheorem{example}{Example}

\usepackage{algorithm}
\usepackage{algpseudocode}

\newcommand{\tmop}[1]{\ensuremath{\operatorname{#1}}}

\usepackage{lscape}
\usepackage{beramono}
\usepackage{listings}
\usepackage[usenames,dvipsnames]{xcolor}
\usepackage{mathtools}
\usepackage{multicol}

\newcommand{\bx}{\mathbf{x}}

\newcommand{\bmu}{\boldsymbol{\mu}}

\lstdefinelanguage{Julia}%
  {morekeywords={abstract,break,case,catch,const,continue,do,else,elseif,%
      end,export,false,for,immutable,import,importall,if,in,%
      macro,module,otherwise,quote,return,switch,true,try,type,typealias,%
      using,while},%
   sensitive=true,%
   alsoother={\$},%
   morecomment=[l]\#,%
   morecomment=[n]{\#=}{=\#},%
   morestring=[s]{"}{"},%
   morestring=[m]{'}{'},%
}[keywords,comments,strings]%

\begin{document}
\begin{frontmatter}

\title{Some Computational Tools for Solving a Selection of Problems in Control Theory} 


\author[First]{Alexander Demin} 
\author[Second]{Christina Katsamaki} 
\author[Third]{Fabrice Rouillier}

\address[First]{HSE University, Moscow, Russia (e-mail: asdemin\_2@edu.hse.ru).}
\address[Second]{Sorbonne Universit\'e, Paris Universit\'e,  CNRS (IMJ-PRG), Inria Paris, Paris, France (e-mail: christina.katsamaki@inria.fr)}
\address[Third]{ Sorbonne Universit\'e, Paris Universit\'e,  CNRS (IMJ-PRG), Inria Paris, Paris, France (e-mail: Fabrice.Rouillier@inria.fr)}

\begin{abstract}                

This paper demonstrates how certified computational tools can be used to address various problems in control theory. In particular, we introduce \texttt{PACE.jl}, a Julia package that implements  symbolic elimination techniques, including (among others) discriminant varieties and Rational Univariate Representation, while also supporting multi-precision interval computations. We showcase its applications to key control theory problems, including identification, stability analysis, and optimization, for both parameter-dependent and parameter-free systems.
\end{abstract}

\begin{keyword}
Symbolic Computation and Control, Multivariable Systems, Parameter Estimation.
\end{keyword}

\end{frontmatter}

\section{Introduction}

Control theory fundamentally relies on mathematical principles to analyze dynamical systems, optimize performance, and ensure stability. As systems grow more complex, the need for advanced computational tools with certified guarantees becomes critical. Traditional numerical methods, while efficient, often lack the rigor required in critical applications. Symbolic methods offer exact results but at a higher computational cost. Bridging this gap is essential for developing robust, scalable control algorithms.

This paper presents a suite of certified computational tools that can assist in tackling different challenging problems in control theory, including parameter identification, stability analysis, and optimization. To this end, we introduce \textbf{PACE.jl}{\footnote{\url{https://pace.gitlabpages.inria.fr/pace.jl}}}, a Julia package that implements symbolic elimination techniques such as discriminant varieties and Rational Univariate Representation, while also providing support for multi-precision interval computations.
The choice of Julia leverages its high performance and vibrant scientific ecosystem.
All experiments are reproducible and accompanied by notebooks\footnote{\url{https://pace.gitlabpages.inria.fr/pace.jl/applications-control}} with code and data.


\section{Key computable objects}
\label{sec:computational_tools}

\noindent {\bf Polynomial systems with rational coefficients.} Given a system of polynomial equations $\mathcal{S}$ with rational
coefficients (floating point numbers are mathematically rational numbers),
depending on its properties, it can be ``solved exactly'' in several ways.
For example, the roots of $P (X) =
\sum_{i = 0}^n a_i X^i, a_i \in \mathbb{Q}$, can be ``exactly'' represented by: 
\begin{itemize}
  \item an \textit{isolating interval} $\left] \frac{n_l}{d_l}, \frac{n_r}{d_r} \right[$
  containing a unique real solution of $P (X) = 0$ with the convention that
  $\left] \frac{b}{c}, \frac{b }{c} \right[$ is exactly the rational number
  $\frac{b}{c}$,
  \item $\overline{P}$, the square-free part of $P$, that is
  ${\overline{P}}=\frac{P}{\gcd(P, P')}$.
\end{itemize}
Such a representation allows, for example, to provide a numerical
approximation of the roots with an arbitrary precision set by the user,
$\overline{P}$ taking opposite signs at the left and at the right of any
isolating interval. Using simple algorithms one can also compare, add, multiply roots.

When a system $\mathcal{S}$ depends on several variables $X_1, \dots, X_n$, but has a finite number of
complex roots, which can be detected at an intermediate step in the
computations, one can simplify the study by computing a, so called, \textit{Rational
Univariate Representation} (RUR) of the roots \citep{rouillier:inria-00098872}, that consists of a linear form with rational
coefficients $t = \sum_{i = 0}^n a_i X_i$ that is injective on the set of
complex roots and $n + 1$ univariate polynomials with rational coefficients in
a new independent variable: $f_t (T), f_1 (T), \ldots, f_n (T)$. If $V
(\mathcal{S}) \subset \mathbb{C}^n$ is the set of complex roots of
$\mathcal{S}$ and $V (f_t) \subset \mathbb{C}$ the set of complex roots of
$f_t$, the RUR associated to $t$ defines a bijection between $V (\mathcal{S})$
and $V (f_t)$ that preserves the real roots and the multiplicities:

\begin{center}
  $\begin{array}{ccc}
    V (\mathcal{S}) & \longrightarrow & V (f_t)\\
    \alpha = (\alpha_1, \ldots, \alpha_n) & \rightarrow & t (\alpha) = \sum_{i
    = 1}^n a_i \alpha_i\\
    \left( \frac{f_1 (\beta)}{\overline{f_t}^{'} (\beta)}, \ldots, \frac{f_n
    (\beta)}{\overline{f_t}^{'} (\beta)} \right) & \leftarrow & \beta
  \end{array}$
\end{center}

If the real roots of $f_t$ have been isolated by means of intervals with rational bounds $\left\{ \left]
\frac{n_{i, l}}{d_{i, l}}, \frac{n_{i, r}}{d_{i, r}} \right[, i = 1 \ldots l
\right\}$, one can then use multi-precision interval arithmetic in order to compute boxes $\frac{f_1 \left(
\left] \frac{n_{i, l}}{d_{i, l}}, \frac{n_{i, r}}{d_{i, r}} \right[
\right)}{\overline{f_t}^{'} \left( \left] \frac{n_{i, l}}{d_{i, l}},
\frac{n_{i, r}}{d_{i, r}} \right[ \right)} \times \ldots \times \frac{f_n
\left( \left] \frac{n_{i, l}}{d_{i, l}}, \frac{n_{i, r}}{d_{i, r}} \right[
\right)}{\overline{f_t}^{'} \left( \left] \frac{n_{i, l}}{d_{i, l}},
\frac{n_{i, r}}{d_{i, r}} \right[ \right)}$ with rational bounds that isolate
the real roots of $V (\mathcal{S})$ and that can be refined up to an arbitrary
precision fixed by the user.
Given a root represented by an isolating interval, a common method for refining 
it—whether to achieve higher precision or to ensure certain properties, such as 
sign invariance when evaluating another nonzero polynomial at the root—involves 
applying the interval Newton's method \citep{revol:inria-00072253}. This iterative 
approach refines the initial interval by computing Newton steps in an interval form 
and intersecting the result with the previous interval at each iteration. This process 
progressively narrows the isolating interval, improving the accuracy of the root representation.

\noindent {\bf Parametric polynomial systems.} Given a system of polynomial equations depending on parameters, the notion of {\em solving} a system is not straightforward.

For problems depending on one variable and several parameters, 
for example given $P_1, P_2 \in \mathbb{Q} [U_1,\ldots,U_l][X]$, one can perform elimination by
considering the {\em subresultant sequence}~(see  \cite{10.1145/321371.321381}),  $\{\tmop{Sres}_d, \ldots, \tmop{Sres}_0\}$, of $P_1, P_2$ with respect to the variable $X$. This sequence has the following properties:

$\bullet$ $\tmop{Sres}_i \in \mathbb{Q}[U_1,\ldots,U_l][X]$ has degree $i$ in $X$ or is null.
  
$\bullet$ $\tmop{Sres}_0 \in \mathbb{Q} [U_1,\ldots,U_l]$ is the resultant of $P_1, P_2$.

$\bullet$ Given $\alpha=(\alpha_1,\ldots,\alpha_l)\in \mathbb{C}^l$, the sequence $\{\tmop{Sres}_d(\alpha,X),$ $ \ldots, \tmop{Sres}_0(\alpha,X)\}$ is the remainder sequence that appears in the Euclidean algorithm applied to $P_1(\alpha,X)$ and $P_2(\alpha,X)$, up to multiplication by some scalars. In particular,  if $i_{\alpha}$ is the smallest
  index such that $\tmop{Sres}_{i_{\alpha}} (\alpha , X) \neq 0$, then \
  $\tmop{Sres}_{i_{\alpha}} (\alpha, X)$ is proportional to $\gcd (P_1 (\alpha, X), P_2
  (\alpha, X))$. Moreover, if $(\alpha,\beta)$ is a root of $P_1=P_2=0$, then $\alpha$ is a root of $Sres_0$.

One can also compute the \textit{Sturm-Habicht sequence} of $P_1$ and $P_2$, that is essentially the subresultant sequence of $P_1$ and the remainder obtained from the Euclidean division of $P_1'P_2$  by $P_1$, with systematic sign modifications \citep{10.1007/978-3-7091-9459-1_14}. If we forget the parameters $U_1, \dots, U_l$, sign variations in the Sturm-Habicht sequence serve in computing the difference of cardinalities 
$\sharp \left( \left\{ x \in \mathbb{R} \| P_1(x) = 0, P_2(x) > 0 \right\} \right)
- \sharp \left( \left\{ x \in \mathbb{R} \| P_1(x) = 0, P_2(x) < 0 \right\} \right)$ \cite[Thm.~4.1]{10.1007/978-3-7091-9459-1_14}.
For $P_2=1$, this gives the number of real roots of $P_1$. The good specialization properties of Sturm-Habicht sequences allow for real root counting also when dealing with parametric polynomials. Given that $P_1, P_2 \in \mathbb{Q} [U_1, \dots, U_l][X]$, 
the sequence can first be computed symbolically for general parameter values $U_1, \dots, U_l$. Once the sequence 
is established, it can then be specialized by substituting specific values $u_1, \dots, u_l$ for the parameters, 
after which the sign variations of the resulting Sturm-Habicht sequence determine the number of real roots of the specialized polynomial.

Given $P \in \mathbb{Q} [U_1,\ldots,U_l][X]$, the number of complex roots of $P$ varies whenever the parameters $U_1,\ldots,U_l$ cancel the leading coefficient of $P$ wrt $X$ (number of complex roots counted with multiplicities decreases) or when the discriminant of $P$ wrt $X$, which is a polynomial in $\in \mathbb{Q} [U_1,\ldots,U_l]$ cancels (number of distinct roots decreases). 
Said differently, over an open ball in the parameter's space that does not meet the resultant of $P$ and $\frac{\partial P}{\partial X}$, the roots of $P$ define an analytic cover of the ball (the implicit function theorem applies on each of the leafs).

This notion has been generalized in \cite{lazard:hal-01148721} for systems depending on parameters. Suppose that $P_1,\ldots,P_s \in \mathbb{Q} [U_1,\ldots,U_l][X_1,\ldots ,X_n]$ for a zero-dimensional system for almost all parameter values (the $U_1,\ldots,U_l$), the authors define the \textit{discriminant variety} of $P_1,\ldots,P_s$, which can be computed by means of Gröbner bases, as the zero set of some polynomial equations in $U_1,\ldots,U_l$ such that over any ball in the parameter's space that does not meet the discriminant variety, the zero set of the system defines an analytic covering of the ball.

Given a discriminant or a discriminant variety, both defined as a union of zeroes of polynomials depending on the parameters, one has to describe the connected components of its complement, say the regions above which the solutions are mathematically stable (the implicit function theorem can be applied on each leaf). In the real case, we are then interested in the regions where the polynomials defining the discriminant variety are of constant and non null sign, which corresponds to the cells of maximum dimension in a \textit{Cylindrical Algebraic Decomposition} (CAD) \citep{10.1007/3-540-07407-4_17} of the parameters' space adapted to the polynomials defining the discriminant variety.

The entire parameter's space can then be viewed as the union of these cells and
of the discriminant variety. In practice, such a decomposition can be described
by a set of sets of polynomials and a set of test points, given $\tmop{DV}$,
the set of polynomials defining the discriminant variety :

$\bullet$  $\tmop{Po}_i \in \mathbb{Q} [U_1, \dots, U_i]$, $i=1,\dots, l$:
  \begin{itemize}
    \item[-] $\tmop{Po}_l = \{p \in \tmop{DV}\} \subset \mathbb{Q} [U_1, \ldots,
    U_l]$
    
    \item[-] $Po_{k - 1} = \{\tmop{Sres}_0 (P, \frac{\partial P}{\partial U_k}),
    \tmop{Sres}_0 (P, Q), P, Q \in P_k \} \subset \mathbb{Q} [U_1, \ldots, U_{k -
    1}]$, $k = l \ldots 2$
  \end{itemize}
$\bullet$ $\tmop{Pt}_1 \in \mathbb{Q}, \ldots \tmop{Pt}_j \in \mathbb{Q}^j,
  \ldots, \tmop{Pt}_l \in \mathbb{Q}^l$ :
  \begin{itemize}
    \item[-] $\tmop{Pt}_1 = \{a_{1, 0}, \ldots a_{1, d_1} \}$ where $a_{1, 0},
    \ldots a_{1, d_1} \in \mathbb{Q}$,  $- \infty < a_{1, 0} < \alpha_{1, 1},
    \ldots, \alpha_{1, i} < a_{1, i} < \alpha_{1, i + 1}, \ldots, a_{1, d_i} <
    \alpha_{1, d_i} < \infty$ and $\alpha_{1, 1}, \ldots, \alpha_{1, d_1}$ are
    the real roots of $\prod_{p \in P_1} p (U_1)$.
    
    \item[-] $\tmop{Pt}_j = \{ (b, a_{j, 1}), \ldots, (b, a_{j, d_j}), b \in
    \tmop{Pt}_{j - 1} \}$ where $a_{j, 0}, \ldots a_{j, d_1} \in \mathbb{Q}$, 
    $- \infty < a_{j, 0} < \alpha_{j, 1}, \ldots, \alpha_{j, i} < a_{j, i} <
    \alpha_{j, i + 1}, \ldots, a_{j, d_i} < \alpha_{j, d_i} < \infty$ and
    $\alpha_{j, 1}, \ldots, \alpha_{j, d_1}$ are the real roots of $\prod_{p
    \in P_j} p (b, U_j)$.
  \end{itemize}

Geometrically, a cell in dimension $1$ is an open interval between two roots
of $\prod_{p \in Po_1} p (U_1)$ and a cell $C_k$ in dimension $k$ is a cylinder
over a cell $C_{k - 1}$ in dimension $k - 1$ delimited by the zeroes of
$\prod_{p \in Po_j, b \in C_{k - 1}} p (b, U_j)$. By construction, the zeroes
of $\prod_{p \in Po_j, b \in C_{k - 1}} p (b, U_j)$ define an analytic covering of
$C_{k - 1}$.

\section{\textbf{PACE.jl}: A Julia Toolbox}
\label{sec:pacejl}

\texttt{PACE.jl} is a Julia package that serves as a wrapper and integration point for several specialized packages, while also introducing its own functions. Below, we list the included packages that we have developed:

$\bullet$ \texttt{DiscriminantVariety.jl}\footnote{\url{https://github.com/sumiya11/DiscriminantVariety.jl}}: Discriminant variety computations in polynomial systems \citep{lazard:hal-01148721}. 

$\bullet$ \texttt{LACE.jl}\footnote{\url{https://gitlab.inria.fr/pace/lace.jl}}: Newton's method with interval arithmetic \citep{revol:inria-00072253}.

$\bullet$ \texttt{MPFI.jl}\footnote{\url{https://ckatsama.gitlabpages.inria.fr/mpfi.jl/}}: Wrapper for the MPFI library for multiple-precision interval computations in Julia. See \citep{revol:inria-00544998} for the specifications.

$\bullet$ \texttt{RationalUnivariateRepresentation.jl}\footnote{\url{https://gitlab.inria.fr/newrur/code/-/tree/main/Julia/RationalUnivariateRepresentation.jl}}: Algorithm from \citep{demin2024readingrationalunivariaterepresentations} for the RUR computation of zero-dimensional polynomial systems. 

$\bullet$ \texttt{RS.jl}\footnote{\url{https://pace.gitlabpages.inria.fr/rs.jl}}: Computes isolating intervals for the real roots of a polynomial or a zero-dimensional polynomial system. The package implements the algorithms from \citep{rouillier:inria-00099941} and \citep{kobel:hal-01363955}.

\texttt{PACE.jl} also includes implementations for partial CAD \citep{10.1007/3-540-07407-4_17} and Sturm-Habicht sequence computation.
\texttt{RS.jl} and \texttt{DiscriminantVariety.jl} make use of the Julia package \url{https://github.com/sumiya11/Groebner.jl}{\texttt{Groebner.jl}} \citep{demin2024groebnerjlpackagegrobnerbases}. For polynomial manipulations we rely on \texttt{Nemo.jl} and \texttt{AbstractAlgebra.jl}. The supported architectures include Windows (x86\_64), Linux (x86\_64), and macOS (ARM64). 
\section{Applications in Control Theory}

\subsection{Parameter Identification}
\label{sec:parameter_identification}

Parameter identification in ODE control systems consists in recovery of unknown parameter and initial condition values from measured data. 
Within the framework of differential algebra, ~\citep{bassik:2023} proposes to reduce parameter estimation to solving polynomial systems. The polynomial system is constructed from the ODE's equations and a time-series of the measured output by differentiating the ODE several times. Naturally, solving large nonlinear systems is hard, and this part of the algorithm may become a bottleneck when estimating parameters in large dynamical models. 
The goal of this section is to demonstrate how our implementations may assist in solving such polynomial systems in a certified way.


We begin with the problem formulation from \citep{bassik:2023}. 
%
%
We will use bold letters for tuples of variables.
We consider dynamical models given by ODEs {\color{black}in the state-space form}
\begin{equation}
\label{eq:system}
\begin{cases}
    \mathbf{x}'(t) = \mathbf{f}(\mathbf{x}(t), \boldsymbol{\mu}, \mathbf{u}(t)),\\
    \mathbf{y}(t) = \mathbf{g}(\mathbf{x}(t), \boldsymbol{\mu}, \mathbf{u}(t)),
\end{cases}
\end{equation}
where $t$: time variable, $\mathbf{x}$: state variables, $\mathbf{y}$: (measured) output variables, $\boldsymbol{\mu}$: time-independent parameters, $\mathbf{u}$: (known) control variables, and $\mathbf{f} = (F_1,\ldots,F_n)$ and $\mathbf{g} = (G_1, \ldots, G_m)$: tuples of elements of $\mathbb{C}(\mathbf{x}(t), \boldsymbol{\mu}, \mathbf{u}(t))$. In what follows, we may omit the explicit dependence on time.
%
%
%
To simplify presentation, suppose that there is only one output function denoted by $y = g(\bx, \bmu)$, and that the right-hand side of the ODE is polynomial.

Given an ODE as in \eqref{eq:system} and a time-series of measured data $D = \{(t_1, y(t_1)), \ldots, (t_n, y(t_n))\}$, the task of parameter identification is to estimate the unknown values of parameters $\boldsymbol{\mu}$ and initial conditions $\mathbf{x}(t_0)$ for $t_0 \in \mathbb{R}$.

For brevity, let $\bx_0 \coloneqq \bx(t_0)$. For a sufficiently high order $h$ (which is made precise in \cite{hong:2019}), the ODE is differentiated $h$ times to obtain the equations
\[y = g, ~y' = g', ~\ldots,~ y^{(h)} =g^{(h)}.\]
Assuming we have access to $y$ and its first $h$ derivatives, the equations can be evaluated at time $t_0$ to obtain a polynomial system in $\mathbb{C}[\bmu, \bx_0, \bx_0', \ldots, \bx_0^{(h)}]$. Derivatives are estimated by interpolating (or approximating) the data $D$ as a function of time $\hat{y}(t)$ and differentiating $\hat{y}(t)$ symbolically. The solution $\bmu, \bx_0$ to the original parameter estimation problem can be extracted from the solutions of the polynomial system.

To illustrate the method, we consider the toy ODE system
\[
\left.\begin{aligned}
x' = \mu^2 x,\quad y = x^2 + x,
\end{aligned}\right.
\]
together with the data observed in an experiment: 
\[
D = \{(0.00, 2.00), (0.33, 2.40), (0.67, 2.89), (1.00, 3.49)\}.
\]
We used the initial condition $x_0 = 1.00$ with $t_0 = 0.00$ and $\mu = 0.60$ to simulate the ODE and obtain the data. The goal of parameter estimation is to recover these values.

By differentiating the output twice, we construct a polynomial system:
\begin{equation*}
    \label{eq:poly-sys2}
    \begin{aligned}
        y &= x^2 + x \\
        y' &= 2 x x' + x'\\
        y'' &= 2 x' x' + 2 x x'' + x''\\
    \end{aligned}
    \quad\quad
    \begin{aligned}
        x' &= \mu^2 x\\
        x'' &= \mu^2 x'\\
    \end{aligned}
\end{equation*}
The next step is to compute a function $\hat{y}(t)$ that approximates the data $D$.
For example, rational interpolation yields $\hat{y}(t) = \frac{7.40 t + 54.02}{ 1.52 t^2 - 10.93 t + 27.01}$.

Using the approximation $\hat{y}(t) \approx y(t)$, we evaluate the equations at $t_0 = 0.00$ to obtain a system on $\mu, x_0, x'_0, x''_0$:
\begin{equation*}
    \label{eq:poly-sys3}
    \begin{aligned}
        1.000 &= x_0^2 + x_0 \\
        0.608 &= 2 x_0 x_0' + x_0'\\
        0.227 &= 2 x_0' x_0' + 2 x_0 x_0'' + x_0''\\
    \end{aligned}
    \quad\quad
    \begin{aligned}
        x_0' &= \mu^2 x_0\\
        x_0'' &= \mu^2 x_0'\\
    \end{aligned}
\end{equation*}

A square subsystem has $4$ solutions:
\[(\mu, x_0) = (\pm 0.604, 1.000)\]
\[(\mu, x_0) = (\pm 0.427, -2.000)\]
To find the best fit out of these solutions, different heuristics can be employed. For example, if values are expected to be non-negative, one readily obtains the only positive solution, $(\mu, x_0) = (0.604, 1.000)$. This gives the estimated values for the parameter and the initial condition.

\noindent

\begin{example}
To showcase our tools, we consider a model of NF-kB action~\citep{nfkb}; NF-kB proteins regulate immune response to infection. Our version of the model has 16 states and 5 unknown parameters; we generate synthetic data with 21 data-points.
We used the Julia package {\texttt ParameterEstimation.jl}~\citep{bassik:2023} to produce a polynomial system from the ODE and data. Result is a nonlinear polynomial system in $73$ indeterminates.

In Julia, we load the necessary packages and the definition of the polynomial system:
\begin{lstlisting}
using AbstractAlgebra, PACE, MPFI
system = include("nfkb.txt")
\end{lstlisting}
We then compute a convenient representation of solutions without introducing approximation errors:
\begin{lstlisting}
rur, sep = zdim$\_$parameterization(
            system,
            get$\_$separating$\_$element=true);
\end{lstlisting}
To obtain numerical solutions, we invoke a root-finding routine
from {\texttt PACE.jl} 
specifying desired precision:
%
\begin{lstlisting}
sol = isolate(rur,sep,output$\_$precision=100);
\end{lstlisting}%
Result is 8 solutions to the polynomial system. {\color{black}We simulate the ODE for each candidate solution and pick the best one based on the goodness of fit against available data:}
\vspace{-1.5em}
\begin{multicols}{2}
{\em ground-truth}
\begin{lstlisting}
k$\_$prod   0.4
t1      0.5
t2      0.6
i1      0.2
i1a     0.3
\end{lstlisting}
\columnbreak
{\em our solution}
\begin{lstlisting}
0.400034
0.498517
0.599466
0.199974
0.29607
\end{lstlisting}
\end{multicols}
\vspace{-1.5em}
The relative error of this solution is around 1\%. It is possible to further refine it by using Newton methods
:
\begin{lstlisting}
initial$\_$point = mid.(sol[1])
sol$\_$refined = interval$\_$newton(
  system, initial$\_$point)
\end{lstlisting}
The result is a refined solution with 6 correct decimal digits. Executing this sequence of commands takes several seconds on commodity hardware. {\color{black}This experiment is also available in a runnable notebook at [\url{https://pace.gitlabpages.inria.fr/pace.jl/ode-identification.html}].}
\end{example}


\subsection{Stability Analysis}
\label{sec:stability_analysis}

{\em Multidimensional systems} (or {\em $n$-D systems}) is a class of systems for which the information propagates in $n>1$ dimensions. Within the frequency domain approach, such a system is defined by means of a transfer function $G$ of the following form   
\begin{equation}\label{def:G}
G(z_1,\ldots,z_n)=\frac{N(z_1,\ldots,z_n)}{D(z_1,\ldots,z_n)}, 
\end{equation}
where $D, \, N \in \mathbb{R}[z_1, \ldots, z_n]$ with ${\rm gcd}(D, N)=1$. A fundamental issue in the systems theory is {\em stability analysis}. 

We illustrate how to put into practice, using the \texttt{PACE.jl} Julia package, the strategies from \citep{bouzidi:hal-01951765} and \citep{bouzidi:hal-01366202} for testing the {\em structural stability} of multidimensional discrete linear systems defined by a multivariate rational transfer function. Such a system is said to be {\em structurally stable} if
the denominator $D$ of $G$ is devoid of zeros in the complex unit polydisc
$\mathbb{U}^n = \prod_{k=1}^n{\{z_k \in \mathbb{C} \ | \ |z_k| \leq 1\}}.$
In other words, if $V_{\mathbb{C}}(D)=\{z=(z_1, \ldots, z_n) \in \mathbb{C}^n \; | \; D(z)=0\}$ denotes the hypersurface formed by the complex zeros of $D$, then (\ref{def:G}) is structurally stable if the following condition holds:
\begin{equation}\label{eq:1}
V_{\mathbb{C}}(D) \cap \mathbb{U}^n=\emptyset.
\end{equation}

Thanks to \cite{DeCarlo77}, the problem can be turn into deciding if $n$ univariate polynomials have solutions in the unit $1$-dimensional polydisk and if an $n$-variate hypersurface intersects the $n$-dimensional polydisk, the condition (\ref{eq:1}) being equivalent to: 
$$\left \{
\begin{array}{ll}
D(z_1,1,\ldots,1) \neq 0, & |z_1| \leq 1, \vspace{1mm}\\
D(1,z_2,1,\ldots,1) \neq 0, & |z_{2}| \leq 1, \vspace{1mm}\\ 
\hspace{1.4cm}  \vdots & \hspace{2mm}  \vdots \vspace{1mm}\\
D(1,\ldots,1,z_n) \neq 0, &  |z_n| \leq 1, \vspace{1mm}\\
D(z_1,\ldots,z_n) \neq 0, &  |z_1|=\ldots=|z_n|=1.
\end{array}
\right.$$

The stability condition of a univariate polynomial can be expressed by means of signed subresultant (Sturm-Habicht) sequences \cite{BPR06}[Thm.~9.41], providing thus a simple algorithm for the $n$ first conditions.
Moreover, \cite{bouzidi:hal-01951765} introduce a Moebius change of variables to turn the last (complex) condition into a condition over the reals. Precisely, the numerator of $D(\frac{x_1-i}{x_1+i},\ldots ,\frac{x_n-i}{x_n+i})$ can be written as
${\cal R}(x_1,\ldots,x_n)+i \cdot {\cal I}(x_1,\ldots,x_n)$, where ${\cal R},{\cal I}\in \mathbb{R}[x_1, \ldots, x_n]$ and the solutions of 
$D=0$ such that $|z_i|=1,z_i\neq 1,i=1,\ldots, n$ are in bijection with the solutions in 
${\cal S}=\{x \in \mathbb{R}^n | {\cal R}=0,{\cal I}=0\}$.

When $n=2$, the system $\{{\cal R}=0,{\cal I}=0\}$ has a finite number of complex roots (zero-dimensional) so that it suffices to isolate its real roots to conclude. 
When $n>2$  the system is positive dimensional (infinite number of complex roots) and an additional processing is required. For example, one can consider the critical points of the distance of the variety ${\cal S}$ to some arbitrary point as in \citep{AUBRY2002543} in order to compute zero-dimensional systems with points on each of the semi-algebraically connected components of ${\cal S}$.

When $n=2$ and $D$ depends on parameters $U1, \dots, U_l$, we follow \citep{bouzidi:hal-01951765}. The above system $\{{\cal R}=0,{\cal I}=0\}$ has coefficients in $\mathbb{Q}[U_1,\ldots,U_l]$. Then, the question of stability resumes in finding the (real) parameter values for which the system has (or not) real roots. As described in \citep{bouzidi:hal-01951765}, the method then consists in computing $DV$, the discriminant variety of the system wrt the projection onto the parameters' space and then to compute the cells of maximal dimension in a CAD of  $\mathbb{R}^l$ adapted to $DV$: above each of these cells, the system is always stable or always unstable, and this property can be addressed for each cell by solving the zero-dimensional system at each sample point of the CAD.

\begin{example}
\label{ex:stab}
For $n=l=2$, we implement the method in \texttt{PACE.jl}.
We begin by defining the polynomial ring:

\begin{lstlisting}
using PACE, Nemo

Rc, (z1, z2, u1, u2) = polynomial$\_$ring(
        Nemo.ComplexField(), 
        [:z1, :z2, :u1, :u2])
\end{lstlisting}

Next, we define the polynomial \( D \) (taken from \citep{LI20212597}):

\begin{lstlisting}
D = u2*z1*z2-u1*z2-u2*z1+z1*z2-z1+1
\end{lstlisting}

Now, we compute the stability regions using the function:

\begin{lstlisting}
stable, unstable = stability$\_$parametric(D)
\end{lstlisting}

This function samples a rational point in each cell of maximal dimension of the CAD of $\mathbb{R}^2$ adapted to the discriminant variety and the conditions resulting from the Lienard-Chipart criterion  \citep{BPR06}[Thm.~9.41] for the stability of the univariate polynomials $D(z1, 1), D(1, z2)$.
The output is:
\[
\text{Stable: } [0, -1]
\]
\[
\text{Unstable: } \left[
\begin{array}{c}
\left[-\frac{3}{2}, -\frac{7}{4}\right], \left[-\frac{3}{2}, -1\right], \left[-\frac{3}{2}, -\frac{1}{4}\right], \\
\left[0, -2\right], \left[0, 0\right], \left[\frac{3}{2}, -\frac{11}{4}\right], \left[\frac{3}{2}, -1\right], \left[\frac{3}{2}, \frac{3}{4}\right]
\end{array}
\right]
\]

\begin{figure}[t]
    \centering
    \includegraphics[scale=0.3]{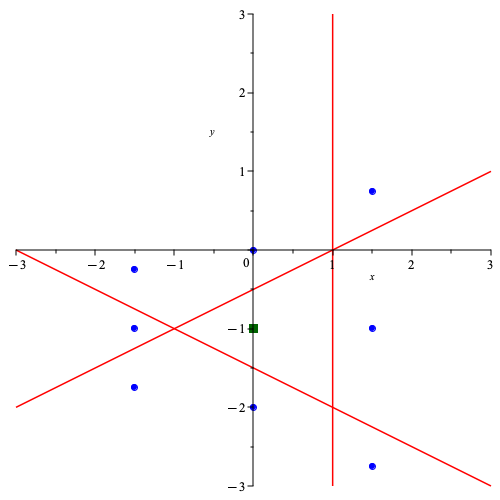}
    \caption{Points sampled in each 2d-cell of the decomposition in Example~2. In blue are the points corresponding to unstable regions and in green to stable.}
    \label{fig:stability}
\end{figure}

It consists of a rational point for each parameter region where the system remains stable or unstable. 
Here, there is only one stable region with representative point $(0,-1)$. These regions are visualized in Figure~\ref{fig:stability}, 
where green points indicate stability and blue points indicate instability. The example can be found and reproduced in [\url{https://pace.gitlabpages.inria.fr/pace.jl/parametric_stability.html}].

\end{example}

\subsection{Optimal Control: Computation of the $H_\infty$
norm}
\label{sec:optimal_control}

In control theory, robustness is a central concept, that is often analyzed through the $L_\infty$-norm, which quantifies the worst-case effects of disturbances and modeling errors. The \textit{$H_\infty$-optimal control theory}, introduced by \cite{1102603}, extends robustness analysis by characterizing stability and performance in terms of the $L_\infty$-norm of a system's transfer matrix restricted to the imaginary axis. However, unlike the $L_2$-norm, the $L_\infty$-norm lacks simple closed-form expressions, requiring numerical or symbolic computation \citep{Tannenbaum}.

The computation of the $L_\infty$-norm of finite-dimensional time-invariant linear systems can be reformulated as finding the maximal $y$-projection of real solutions to a zero-dimensional bivariate polynomial system \citep{Bouzidi_2021}. Before explaining how we can carry out this computation with \texttt{PACE.jl}, we will first describe the problem.

We denote the algebra of real rational functions on the imaginary axis $i \mathbb{R}$ which are proper and without poles on $i \mathbb{R}$ by
$
RL_\infty := \Bigg\{ \frac{n(i \omega)}{d(i \omega)} \mid n,d \in \mathbb{R}[i \omega], \ \gcd(n,d) = 1,  \deg_\omega(n) \leq \deg_\omega(d),  V_{i \mathbb{R}}(\langle d \rangle) = \emptyset \Bigg\}.
$
For a transfer function $G \in \mathbb{R}(s)^{u \times v}$ such that $G_{i \mathbb{R}} \in RL^u_{\infty}$, let  $\gamma > 0$ and
$
\Phi_\gamma(s) := \gamma^2 I_v - G^T(-s)G(s),
$
where $G^T$ is the transpose of $G$. 
From \citep{KANNO2006697}, we have that $\gamma > \| G \|_{\infty}$ if and only if $\gamma > \sigma (G(i \infty))$ 
and $\det(\Phi_\gamma(i \omega)) \neq 0$ 
for all $\omega \in \mathbb{R}$. Note that $\sigma(G(i \infty))$ corresponds to the largest singular value of $G(i \infty)$.

Then, we can write $\det(\Phi_\gamma(i \omega)) = \frac{n(\omega, \gamma)}{d(\omega)},$
where $n \in \mathbb{R}[\omega, \gamma]$ and $d \in \mathbb{R}[\omega]$ are coprime. Since $G$ has no poles on the imaginary axis, $d(\omega)$ does not vanish on $\mathbb{R}$. 
Hence, to compute the maximal singular value of $G(i \omega)$, and in turn the $L_\infty$-norm of $G$, we have to compute the maximal real value 
$\gamma$ satisfying that a real value $\omega$ exists such that $\det(\Phi_\gamma(i \omega))$ vanishes.
In other words, we are led to study the $\gamma$-extremal points of the real plane algebraic curve given by $n(\omega, \gamma) = 0$.
We denote by \( \tilde{n} \in \mathbb{R}[\omega, y] \) the square-free part of \( n \). The \( L_\infty \)-norm of \( G \) is then given by:
\[
\|G\|_\infty = \max \left\{ 
\pi_{\gamma} \left( V_{\mathbb{R}} \left( \tilde{n}, \frac{\partial \tilde{n}}{\partial \omega} \right) \right)
\cup 
V_{\mathbb{R}} \left( \langle \mathrm{lc}_\omega(\tilde{n}) \rangle \right)
\right\},
\]
where $\pi_y$ gives the $y$-projection and $\mathrm{lc}_\omega$ the leading coefficient when considered as a polynomial in $\omega$.



In \citep{Bouzidi_2021}, several certified symbolic methods were proposed for this computation. We chose to implement the {\em Sturm-Habicht} method, introduced as a third approach. It exploits the fact that the curve  given by \(\tilde{n} \in (\mathbb{Z}[\gamma])[\omega]\) is bounded in the 
$\gamma$-direction by the value being sought and relies on real root counting based on the sign variations of the Sturm-Habicht sequence. 
We compute the Sturm-Habicht sequence \(\{\mathrm{StHa}_j(\tilde{n}, 1)\}_{j=0,\ldots,\deg_\omega(\tilde{n})}\) of $\tilde{n}$ and $1$ and the real roots of \(\mathrm{StHa}_0\). Among them, we keep the larger one that is either a root of $lc(\tilde{n})$ or that makes the sign  variations  in the sequence of principal Sturm-Habicht coeﬃcients positive \citep[Def.~2.2, 3.2]{10.1007/978-3-7091-9459-1_14}.






\begin{example}
\texttt{PACE.jl} wraps the computation of the $H_\infty$-norm in the function \texttt{Hinf\_StHa()}.
We begin by importing the necessary Julia packages:
\begin{lstlisting}
using PACE, MPFI, Nemo, AbstractAlgebra

\end{lstlisting}

Then, we define the rational function field and the transfer matrix:

\begin{lstlisting}
S1,(y,s) = rational$\_$function$\_$field(
    AbstractAlgebra.QQ,[:y, :s])

G=[s//(s+1)-s//(s+1);-s//(s+1)1//(s+1)]
\end{lstlisting}
By calling \texttt{Hinf$\_$StHa(G)}, we obtain: $[1.59375, 1.625]$
The output is an interval containing the value of the \( H_{\infty} \) norm. Since a working precision was not explicitly specified, the function chose the minimum precision necessary to isolate the roots. If we wish to specify a binary precision for the computation, we can call \texttt{Hinf\_StHa()} with the \texttt{starting\_precision} argument. For example:

\begin{lstlisting}
Hinf$\_$StHa(P; starting$\_$precision=Int32(10))
\end{lstlisting}

The output is then: $[1.617919921875, 1.6180419921875]$.
This example can be found and reproduced in [\url{https://pace.gitlabpages.inria.fr/pace.jl/hinf_notebook.html}].





\end{example}

\section{Conclusion and Future Work}
\label{sec:conclusion}

In this paper, we introduced \texttt{PACE.jl}, a comprehensive Julia package that implements computational tools for solving various problems in control theory.
We illustrated the applicability of our tools through three examples in control theory, where, by using certified algorithms, we produced results that are guaranteed to be correct. 
Future work will focus on extending the package's functionality to handle more general classes of systems and control problems, 
as well as conducting comprehensive performance evaluations. In particular, we plan to implement methods for parametric 
$H_\infty$-norm computation using discriminant varieties, as proposed in \citep{Quadrat_Rouillier_Younes_2024}. 

\bibliography{ifacconf} 
\end{document}